\documentclass[twocolumn,english,aps,prl,amsmath,amssymb,showpacs]{revtex4}
\usepackage[T1]{fontenc}
\usepackage[latin9]{inputenc}
\usepackage{amsmath}
\usepackage{graphicx}
\usepackage{amssymb}
\usepackage{esint}

\makeatletter
\@ifundefined{textcolor}{}
{%
 \definecolor{BLACK}{gray}{0}
 \definecolor{WHITE}{gray}{1}
 \definecolor{RED}{rgb}{1,0,0}
 \definecolor{GREEN}{rgb}{0,1,0}
 \definecolor{BLUE}{rgb}{0,0,1}
 \definecolor{CYAN}{cmyk}{1,0,0,0}
 \definecolor{MAGENTA}{cmyk}{0,1,0,0}
 \definecolor{YELLOW}{cmyk}{0,0,1,0}
 }

\@ifundefined{definecolor}
 {\usepackage{color}}{}
\makeatother

\makeatother

\usepackage{babel}

\begin{document}

\preprint{This line only printed with preprint option}

\title{A quantum motor: directed wavepacket motion in an optical lattice}

\author{Quentin Thommen}

\author{Jean Claude Garreau }

\author{V{\'e}ronique Zehnl{\'e}}

\affiliation{Laboratoire de Physique des Lasers, Atomes et Mol{\'e}cules, Universit{\'e}
Lille 1 Sciences et Technologies, CNRS; F-59655 Villeneuve d'Ascq
Cedex, France}

\homepage{http://www.phlam.univ-lille1.fr/atfr/cq}
\begin{abstract}
We propose a method for arbitrary manipulations of a quantum wavepacket
in an optical lattice by a suitable modulation of the lattice amplitude.
A theoretical model allows to determine the modulation corresponding
to a given wavepacket motion, so that arbitrary atomic trajectories
can be generated. The method is immediately usable in state of the
art experiments. 
\end{abstract}


\maketitle
The fine manipulation of wavepackets is a fundamental requirement
in a large number of fields. Particularly important examples are quantum
transport \cite{Inguscio:InstabilityBEC:PRL04,Ferrari:QuantumTransport:NP09,Weitz:HamiltRatchet:S09,Abduallaev:QuantumTransport:PRL10,Zhang:QuantumTransport:PRA10},
{}``quantum simulators'' that aim to reproduce solid state models
\cite{BlochI:MottTransition:N02,Bouyer:AndersonBEC:N08,Inguscio:AndersonBEC:N08,Inguscio:AubryAndreInteractions:NP10,AP:Anderson:PRL08,AP:AndersonCritic:PRL10,Naegerl:SuperBlochOsc:PRL10},
quantum information \cite{MonroeWineland:SchrCat:S96,Monroe:QIReview:N02,Shapiro:WavepacketsQI:PRL04,Lengwenus:CohTransport:PRL10}
and quantum metrology \cite{Kimble:QuantumStateMetrol:S08}. The possibility
of trapping very cold atoms by light was a decisive step leading to
enormous progress in this field. Tailored optical potentials, created
by multiple interfering beams interacting with cold atoms, allow to
trap and to guide atomic wavepackets for long times (compared to the
dynamics of the atom external degrees of freedom) and distances (compared
to the de Broglie wavelength). Moreover, using far off-resonance beams
reduces decoherence effects to negligible levels. With such techniques
old problems of quantum dynamics have been experimentally studied,
as the elusive Bloch oscillations \cite{BenDahan:BlochOsc:PRL96,Raizen:WSOptPot:PRL96}
or quantum chaos \cite{Raizen:QKRFirst:PRL95,AP:Reversibility:PRL05,AP:Bicolor:PRL00,Summy:SubFQRes:PRL10,Monteiro:DirectedMotionDKR:PRL07,Korsch:BlochOscBEC:PRE05,Sadgrove:TrasportQR:NJP09}.
In the emerging, and rapidly developing, field of quantum information,
controlled motion in optical lattices provides ways to manipulate
\emph{q-bits} \cite{Zoller:ControlledCollision:ANP05,Meschede:AtomSortingMachine:N06}.
The next step in the development of such techniques is the shaping
and displacing of wavepackets at will, and this is the problem addressed
in the present work. We consider the motion of a wavepacket in a driven
two-dimensional (2D) optical lattice (the generalization to the 3D
case is straightforward). By carefully engineering the temporal driving
of the optical potential, we demonstrate a way to coherently impinge
to an atom in an lattice almost any kind of motion, including coherent
rotations of its wavepacket. Tailored wavepacket motions have been
previously studied both theoretically and experimentally by using
Bloch oscillations \cite{Korsch:BlochOsc2D:NPJ04,Naegerl:SuperBlochOsc:PRL10},
but in this case the amplitude of the motion is directly limited by
the spatial extension of the wavepacket coherence. Controlled motion
in modulated lattices has also been observed experimentally in \cite{Ferrari:QuantumTransport:NP09,Naegerl:SuperBlochOsc:PRL10}.
Our work shows how a suitably {}``overmodulated'' driving allows
to displace a wavepacket along paths of arbitrary shape with controlled
(and even engineered) dispersion.

Let us first consider the quantum dynamics of an atom in a 1D tilted
(or {}``washboard'') potential formed by a sinusoidal potential
superposed to a constant force $F_{x}$, or a linear potential $F_{x}x$.
We use normalized variables such that lengths are measured in units
of the lattice step $d$ ($=\lambda_{L}/2$, $\lambda_{L}$ $=2\pi/k_{L}$
being the laser wavelength), energy in units of the recoil energy
$E_{R}=\hbar^{2}k_{L}^{2}/(2M)$ where $M$ is the atom mass, and
time is measured in units of $\hbar/E_{R}$. The corresponding Hamiltonian
is \begin{equation}
H_{x}=-\frac{1}{2m^{*}}\frac{\partial^{2}}{\partial x^{2}}+\left[V_{x}+A_{x}(t)\right]\cos(2\pi x)+F_{x}x,\label{eq:H1D}\end{equation}
 where $m^{*}=\pi^{2}/2$ is the mass in normalized units, $F_{x}$
is the constant force measured in units of $E_{R}/d$, $\hbar=1$,
and $V_{x}$ is the lattice amplitude, to which a time-dependent component
$A_{x}(t)$ can be added. Tilted optical lattices have been experimentally
realized by many groups \cite{BenDahan:BlochOsc:PRL96,Raizen:WSOptPot:PRL96,Arimondo:LandauZener:PRL09}.

\begin{figure*}[t]

\begin{centering}
\includegraphics[clip,width=12cm]{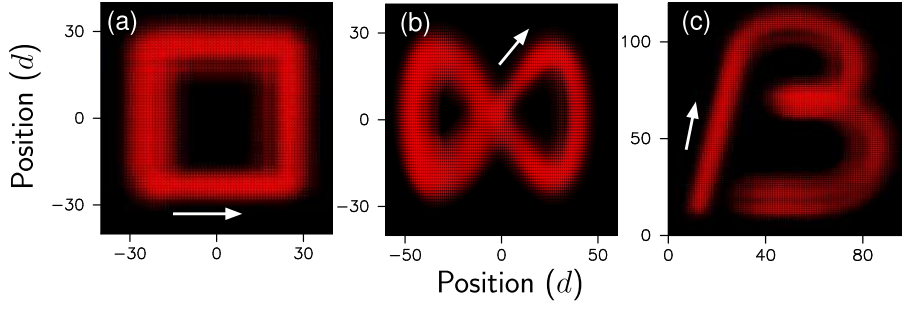} 
\par\end{centering}

\caption{\label{fig:atom_trajectories}(Color online) Arbitrary wavepacket
motion in two dimensions. The plots display the probability of presence
$\left|\Psi_{xy}(t)\right|^{2}$ integrated on time, and the arrows
indicate the sens of the motion. (a) Square path formed of four straight
lines. (b) $\infty$-path, obtained by choosing $\alpha_{x}=\cos\left(\Omega t\right)$,
$\alpha_{y}=\cos\left(2\Omega t\right)$, $\beta_{x}=\beta_{y}=0$.
Plot (c) is made is made by combining straight lines and arcs. Parameters
of the potential are $V_{x,y}=2.5$ and $F_{x,y}=0.2$ (note the different
scales in each plot).}

\end{figure*}

Dynamics in a tilted lattice is conveniently described by \emph{Wannier-Stark}
(WS) states \cite{Wannier:WS:PR60,Raizen:WSOptPot:PRL96,AP:WannierStark:PRA02,Nienhuis:CoherentDyn:PRA01,Korsh:WSOptLatt:PRA00},
which are the eigenstates\emph{ }of the Hamiltonian of Eq.~(\ref{eq:H1D})
with $A_{x}(t)=0$. In numerical simulations a finite lattice is used,
but if the bounds of the lattice are far from the region of interest
there is no essential difference in the dynamics. The eigenenergies
of the Hamiltonian.~(\ref{eq:H1D}) form a {}``ladder'' structure
separated by the {}``Bloch frequency'' $\omega_{B}\equiv2\pi/T_{B}$$=F_{x}$
($\omega_{B}=F_{x}d/\hbar$ in usual units). Depending on the ratio
$V_{x}/F_{x}$ each well can host more than one WS state, each family
of states then forming its own ladder \cite{AP:WannierStark:PRA02}.
Throughout this paper the parameters of the potential and the initial
conditions are chosen so that the atomic dynamics is accurately described
by the \emph{lowest-ladder} WS states only. This situation can be
realized experimentally by using cold enough atoms and raising the
optical potential adiabatically \cite{Raizen:WSOptPot:PRL96}. As
WS states are, for the parameters used in the present work, highly
localized on a potential well, we label a WS state by the well index
$n$ in which it is centered $\varphi_{n}(x)$. These states are invariant
under a translation of a integer number $n$ of lattice steps, provided
the associated energy is also shifted by $n\omega_{B}$, that is \begin{eqnarray}
\varphi_{n}(x) & = & \varphi_{0}(x-n)\label{eq:spatialtranslation}\\
E_{n} & = & E_{0}+n\omega_{B}\end{eqnarray}
 (in what follows we set $E_{0}=0$). It is well known that a wavepacket
submitted to the Hamiltonian of Eq.~(\ref{eq:H1D}) with $A_{x}(t)=0$
has an oscillatory behavior, called Bloch oscillation, of period $T_{B}.$
Adding a time-dependent potential modulated at (or around) the frequency
$\omega_{B}$ is thus a good way to create a resonant response in
the dynamics.

We can expand the atomic wave function $\psi_{x}(t)$ over the WS
states (of the first ladder):\begin{equation}
\psi_{x}(t)=\sum_{n}c_{n}(t)e^{i\phi_{n}(t)}\varphi_{n}(x),\label{eq:WSdevelop}\end{equation}
 where the phase $\phi_{n}(t)$ is conveniently defined by:\begin{equation}
\phi_{n}(t)=-n\omega_{B}t-V_{x}M_{0}\intop_{0}^{t}A_{x}(t^{\prime})dt^{\prime}\label{eq:Phi_n}\end{equation}
 and $M_{0}$ is the $p=0$ value of the coupling parameter $M_{p}$:

\begin{equation}
M_{p}\equiv\left\langle \varphi_{i}\right|\cos(2\pi x)\left|\varphi_{i+p}\right\rangle =\left\langle \varphi_{0}\right|\cos(2\pi x)\left|\varphi_{p}\right\rangle .\label{eq:M_p}\end{equation}
 {[}the last identity is a consequence of Eq.~(\ref{eq:spatialtranslation}){]}.
Bringing Eq.~(\ref{eq:WSdevelop}) into the Schrödinger equation
associated to the Hamiltonian Eq.~(\ref{eq:H1D}) one obtains the
following set of equations:\begin{equation}
\dot{c}_{n}(t)=-iA_{x}\left(t\right)\sum_{p\neq0}M_{p}c_{p+n}e^{-ip\omega_{B}t}.\label{eq:DynamicalEq1}\end{equation}
Eqs.~(\ref{eq:DynamicalEq1}) can be simplified by using the fact
a WS state overlaps significantly only with the WS states associated
with the nearest neighbor sites, that is $M_{p}\approx0$ for $|p|>1$.

\begin{figure*}[t]
\begin{centering}
\includegraphics[width=12cm]{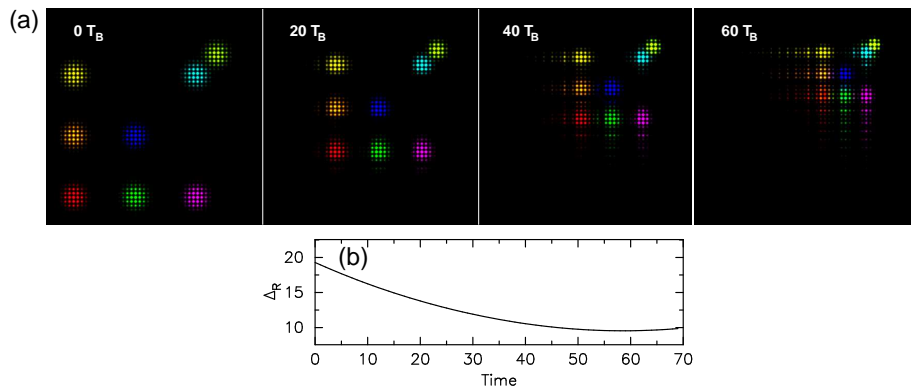} 
\par\end{centering}

\caption{\label{fig:concentration}(Color online) The left plot on the top
row displays a few wavepackets (or different slices of a large wavepacket)
arbitrarily distributed in space. The potential defined by Eq.~(\ref{eq:PotentialCompression})
sets the different wavepackets in motion with a velocity directed
\emph{towards} the origin (situated close to the upright corner) whose
amplitude increases with the distance from the origin. The slices
are thus progressively concentrated around the origin. Plot (b) show
the evolution of the total spatial dispersion with respect to the
origin. Parameters of the potential are $V_{x,y}=2.5$, $F_{x,y}=0.25$
and $k/k_{L}=0.02$.}

\end{figure*}

We now set in Eq.~(\ref{eq:H1D}) the driving as\begin{equation}
A_{x}(t)=\alpha_{x}\left(t\right)\sin\left(\omega_{B}t+\beta_{x}\right).\label{eq:modulation}\end{equation}
where $\alpha_{x}\left(t\right)$ is a slowly varying function ($\left|\alpha_{x}^{-1}d\alpha_{x}/dt\right|\ll\omega_{B}$).
An additional simplification is obtained by neglecting fast oscillating
terms; one then finds from Eq.~(\ref{eq:DynamicalEq1}):

\begin{equation}
\dot{c}_{n}(t)=\frac{M_{1}\alpha_{x}\left(t\right)}{2}\left[c_{n-1}e^{-i\beta_{x}}-c_{n+1}e^{i\beta_{x}}\right]\label{eq:DynamicalEq}\end{equation}
 whose general solution is \begin{equation}
c_{n}(t)=\sum_{p}c_{n+p}(t=0)e^{ip\beta_{x}}J_{p}\left(-M_{1}\int_{0}^{t}\alpha_{x}\left(t^{\prime}\right)dt^{\prime}\right)\label{eq:Sol_an}\end{equation}
 ($J_{p}(x)$ is the Bessel function of the first kind and order $p$)
which is a straightforward extension of a previous result \cite{AP:QuantumDyn:JOBQSO04}.
The mean position of the wavepacket, averaged over a Bloch period
$T_{B}$, is\begin{eqnarray*}
\left\langle x\right\rangle _{t} & = & \left\langle x\right\rangle _{t=0}+\int_{0}^{t}v_{x}\left(t'\right)dt',\end{eqnarray*}
 where \begin{equation}
v_{x}\left(t\right)=M_{1}\alpha_{x}\left(t\right)\mathrm{Re}(\sigma e^{-i\beta_{x}})\label{eq:Vitesse}\end{equation}
 is the instantaneous wavepacket drift velocity and $\sigma\equiv\sum_{p}c_{p}^{*}\left(0\right)c_{p+1}\left(0\right)$
is the initial coherence of the wavepacket. The wavepacket also presents
a diffusion characterized by an instantaneous diffusion coefficient
$D(t)\propto\left[M_{1}\alpha_{x}(t)\mathrm{Im}(\sigma e^{-i\beta_{x}})\right]$.
Hence, the motion \emph{and} the diffusion can be controlled by changing
the temporal driving $\alpha_{x}$ and phase $\beta_{x}$. The diffusion
can be suppressed by setting $\beta_{x}$ so that $\mathrm{Im}(\sigma e^{-i\beta_{x}})=0$;
one then obtains an undeformed translation of the wavepacket with
a velocity given by Eq.~(\ref{eq:Vitesse}), $v_{x}\left(t\right)=\pm M_{1}\alpha_{x}\left(t\right)\left|\sigma\right|$.
Conversely, setting $\beta_{x}$ so that $\mathrm{Re}(\sigma e^{-i\beta_{x}})=0$
leads to a purely diffusive motion with no displacement of the wavepacket's
center of mass.

Consider now a {}``square'' 2D lattice formed by the interference
of two orthogonal pairs of counterpropagating laser beams. It can
be shown that the resulting Hamiltonian is separable \cite{Korsch:BlochOsc2D:NPJ04}:
$H_{xy}=H_{x}+H_{y}$, where $H_{x}$ is given by Eq.~(\ref{eq:H1D})
and $H_{y}$ is the obvious generalization for the $y$ coordinate.
The solution can be written as $\Psi_{xy}(t)=\psi_{x}(t)\psi_{y}(t)$
and the 2D dynamics is obtained simply by solving two identical 1D
Schrödinger equations for each $\psi_{u}(t)$ $(u=x,y)$, with the
corresponding Hamiltonian $H_{u}$. We can thus induce a controlled
2D motion of the wavepacket by choosing suitable lattice modulations
$A_{u}(t)=\!\alpha_{u}(t)\sin(\omega_{B}t+\beta_{u})$.

In order to illustrate the possibilities opened by our method, we
present in Fig.~\ref{fig:atom_trajectories} different trajectories
obtained by numerical integration of the Schrodinger equation with
fixed parameters and an initial wavepacket of gaussian shape \begin{equation}
\Psi_{xy}(0)=\sum_{l,m}\exp(-\frac{l^{2}+m^{2}}{9})\varphi_{l}(x)\varphi_{m}(y).\label{eq:Initial condition}\end{equation}
 Note that this form implies that $\sigma$ is real, the {}``zero
diffusion'' condition is thus fulfilled if $\beta_{x}=\beta_{y}=0$.
Plots (a)-(c) in Fig.~\ref{fig:atom_trajectories} show the square-modulus
of the wavepacket \emph{integrated on time}, and the arrows indicate
the sense of the motion. Plot (a) shows a square trajectory obtained
with $\alpha_{x}=1,\,\,\alpha_{y}=0$, for $0\leq t\le15T_{B}$ ,
$\alpha_{x}=0\,,\,\alpha_{y}=1$ for $15T_{B}\leq t\le30T_{B},$ and
so on. In plot (b), we drive the wavepacket into an $\infty$-shaped
Lissajous curve by setting $\alpha_{x}(t)=\cos\left(\Omega t\right)$,
$\beta_{x}=0$ and $\alpha_{y}(t)=\cos\left(2\Omega t\right)$, $\beta_{y}=0$,
with the overmodulation frequency $\Omega=\omega_{B}/250$. By combining
paths one can generate any type of trajectories in 2D (and, by an
obvious generalization of the above discussion, in 3D): Plot (c) in
Fig.~\ref{fig:atom_trajectories} displays a $\beta$-shaped trajectory,
where even a turning point has been {}``drawn''. In all these numerical
experiments the diffusion is very small, the width of the wavepacket
varies only slightly during the evolution.

New kinds of dynamics can be obtained by using an overmodulation whose
amplitude varies slowly \emph{also in space}, i.e $A_{x}(t)\rightarrow A_{x}(x,t)$
in Eq.~(\ref{eq:H1D}). This can be realized \cite{Arimondo:RatchetOptLat:PRA06,Inguscio:AndersonBEC:N08}
by adding a second laser beam with a different spatial period $k_{L}^{\prime}$,
which produces a spatial modulation of the lattice amplitude corresponding
to the beat note of the two spatial frequencies. The potential is
then \begin{equation}
A_{u}(u,t)=\sin(ku)\left(\sin\omega_{B}t+\beta_{u}\right)\label{eq:PotentialCompression}\end{equation}
 where $k=(k_{L}-k_{L}^{\prime})/k_{L}$. In the limit $|k|\ll1$,
$\sin(ku)\approx ku$ and the velocity of the wavepacket, which is
proportional to the modulation amplitude, varies linearly with the
\emph{position}, as one can deduce from Eq.~(\ref{eq:Vitesse}):
$v_{u}=M_{1}\sigma ku$ (taking $\beta_{u}=0)$. All parts of the
wavepacket, whatever their positions, will move towards the origin
$(0,0)$, and the closer of the origin it is, the smallest its velocity.
This effect can thus be used to concentrate a large wavepacket, or
various wavepackets distributed at arbitrary positions, on a few potential
wells. Fig.~\ref{fig:concentration}(a) shows a numerical experiment
in which such a concentration is realized. Each colored spot represents
a wavepacket, or a part of a large wavepacket. As times goes by, the
different spots are seen to converge to the same position. Plot (b)
in Fig.~\ref{fig:concentration} shows the evolution of the total
spatial dispersion ($\left\langle \Delta r^{2}\right\rangle ^{1/2}$).
This unavoidable dispersion limits the maximum time during which the
overmodulation can be applied, and thus the maximum density which
can be obtained.

\begin{figure*}[t]
\begin{centering}
\includegraphics[width=12cm]{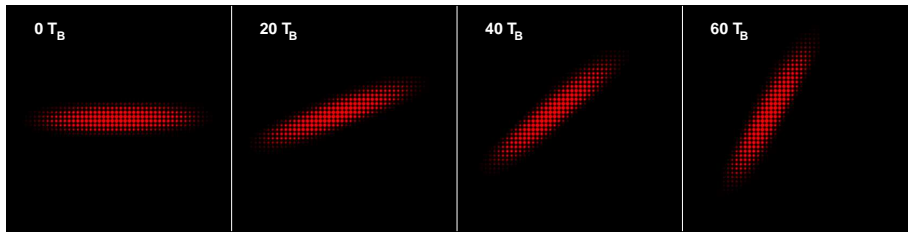} 
\par\end{centering}

\caption{\label{fig:Rotation}(Color online) Rotation of the wavepacket, obtained
by using combined spatial and temporal overmodulations (cf. text).}

\end{figure*}

One can also combine temporal and spatial modulations to produce a
\emph{rotation} of a wavepacket. Consider a cigar-shape wavepacket
as shown in Fig.~\ref{fig:Rotation}(a). Inducing a uniform rotation
around the origin $(0,0)$ means to give a slice at position $\boldsymbol{r}$
a local velocity $\boldsymbol{v}$ perpendicular to $\boldsymbol{r}$.
This can be done by producing a modulation of the form\begin{eqnarray*}
A_{x} & = & -\sin(ky)\sin\left(\omega_{B}t+\beta_{x}\right)\approx-ky\sin\left(\omega_{B}t+\beta_{x}\right)\\
A_{y} & = & \sin(kx)\sin\left(\omega_{B}t+\beta_{y}\right)\approx kx\sin\left(\omega_{B}t+\beta_{y}\right).\end{eqnarray*}
 Figure~\ref{fig:Rotation} shows an example of wavepacket rotation
produced by this technique.

In conclusion, this paper illustrates the almost unlimited power of
overmodulated optical lattices to manipulate atom wavepackets. The
technique is immediately applicable to state-of-art experiments, and
that can play an important role in preparing complex initial state
for fundamental studies of quantum dynamics, but might also useful
for playing with atomic qbits in the realm of quantum information
experiments. Although we have discussed only 2D examples, that are
easier to understand and to display pictorially, all the techniques
illustrated above are straightforwardly generalized to the 3D case. 
\begin{acknowledgments}
Laboratoire de Physique des Lasers, Atomes et Mol{\'e}cules is UMR
8523 du CNRS. Work partially financed by a grant of the Agence Nationale
de la Recherche (MICPAF project). 
\end{acknowledgments}


\begin{thebibliography}{36}
\expandafter\ifx\csname natexlab\endcsname\relax\def\natexlab#1{#1}\fi
\expandafter\ifx\csname bibnamefont\endcsname\relax
  \def\bibnamefont#1{#1}\fi
\expandafter\ifx\csname bibfnamefont\endcsname\relax
  \def\bibfnamefont#1{#1}\fi
\expandafter\ifx\csname citenamefont\endcsname\relax
  \def\citenamefont#1{#1}\fi
\expandafter\ifx\csname url\endcsname\relax
  \def\url#1{\texttt{#1}}\fi
\expandafter\ifx\csname urlprefix\endcsname\relax\def\urlprefix{URL }\fi
\providecommand{\bibinfo}[2]{#2}
\providecommand{\eprint}[2][]{\url{#2}}

\bibitem[{\citenamefont{Fallani et~al.}(2004)\citenamefont{Fallani, De~Sarlo,
  Lye, Modugno, Saers, Fort, and Inguscio}}]{Inguscio:InstabilityBEC:PRL04}
\bibinfo{author}{\bibfnamefont{L.}~\bibnamefont{Fallani}},
  \bibinfo{author}{\bibfnamefont{L.}~\bibnamefont{De~Sarlo}},
  \bibinfo{author}{\bibfnamefont{J.~E.} \bibnamefont{Lye}},
  \bibinfo{author}{\bibfnamefont{M.}~\bibnamefont{Modugno}},
  \bibinfo{author}{\bibfnamefont{R.}~\bibnamefont{Saers}},
  \bibinfo{author}{\bibfnamefont{C.}~\bibnamefont{Fort}}, \bibnamefont{and}
  \bibinfo{author}{\bibfnamefont{M.}~\bibnamefont{Inguscio}},
  \bibinfo{journal}{Phys. Rev. Lett.} \textbf{\bibinfo{volume}{93}},
  \bibinfo{pages}{140406} (\bibinfo{year}{2004}).

\bibitem[{\citenamefont{Alberti et~al.}(2009)\citenamefont{Alberti, Ivanov,
  Tino, and Ferrari}}]{Ferrari:QuantumTransport:NP09}
\bibinfo{author}{\bibfnamefont{A.}~\bibnamefont{Alberti}},
  \bibinfo{author}{\bibfnamefont{V.~V.} \bibnamefont{Ivanov}},
  \bibinfo{author}{\bibfnamefont{G.~M.} \bibnamefont{Tino}}, \bibnamefont{and}
  \bibinfo{author}{\bibfnamefont{G.}~\bibnamefont{Ferrari}},
  \bibinfo{journal}{Nature Phys.} \textbf{\bibinfo{volume}{5}},
  \bibinfo{pages}{547} (\bibinfo{year}{2009}).

\bibitem[{\citenamefont{Salger et~al.}(2009)\citenamefont{Salger, Kling,
  Hecking, Geckeler, Morales-Molina, and Weitz}}]{Weitz:HamiltRatchet:S09}
\bibinfo{author}{\bibfnamefont{T.}~\bibnamefont{Salger}},
  \bibinfo{author}{\bibfnamefont{S.}~\bibnamefont{Kling}},
  \bibinfo{author}{\bibfnamefont{T.}~\bibnamefont{Hecking}},
  \bibinfo{author}{\bibfnamefont{C.}~\bibnamefont{Geckeler}},
  \bibinfo{author}{\bibfnamefont{L.}~\bibnamefont{Morales-Molina}},
  \bibnamefont{and} \bibinfo{author}{\bibfnamefont{M.}~\bibnamefont{Weitz}},
  \bibinfo{journal}{Science} \textbf{\bibinfo{volume}{326}},
  \bibinfo{pages}{1241} (\bibinfo{year}{2009}).

\bibitem[{\citenamefont{Abdullaev and
  Kraenkel}(2000)}]{Abduallaev:QuantumTransport:PRL10}
\bibinfo{author}{\bibfnamefont{F.~K.} \bibnamefont{Abdullaev}}
  \bibnamefont{and} \bibinfo{author}{\bibfnamefont{R.~A.}
  \bibnamefont{Kraenkel}}, \bibinfo{journal}{Phys. Rev. A}
  \textbf{\bibinfo{volume}{62}}, \bibinfo{pages}{023613}
  (\bibinfo{year}{2000}).

\bibitem[{\citenamefont{Zhang and Liu}(2010)}]{Zhang:QuantumTransport:PRA10}
\bibinfo{author}{\bibfnamefont{J.~M.} \bibnamefont{Zhang}} \bibnamefont{and}
  \bibinfo{author}{\bibfnamefont{W.~M.} \bibnamefont{Liu}},
  \bibinfo{journal}{Phys. Rev. A} \textbf{\bibinfo{volume}{82}},
  \bibinfo{pages}{025602} (\bibinfo{year}{2010}).

\bibitem[{\citenamefont{Greiner et~al.}(2002)\citenamefont{Greiner, Mandel,
  Esslinger, H{\"a}nsch, and Bloch}}]{BlochI:MottTransition:N02}
\bibinfo{author}{\bibfnamefont{M.}~\bibnamefont{Greiner}},
  \bibinfo{author}{\bibfnamefont{O.}~\bibnamefont{Mandel}},
  \bibinfo{author}{\bibfnamefont{T.}~\bibnamefont{Esslinger}},
  \bibinfo{author}{\bibfnamefont{T.~W.} \bibnamefont{H{\"a}nsch}},
  \bibnamefont{and} \bibinfo{author}{\bibfnamefont{I.}~\bibnamefont{Bloch}},
  \bibinfo{journal}{Nature (London)} \textbf{\bibinfo{volume}{415}},
  \bibinfo{pages}{39} (\bibinfo{year}{2002}).

\bibitem[{\citenamefont{Billy et~al.}(2008)\citenamefont{Billy, Josse, Zuo,
  Bernard, Hambrecht, Lugan, Cl{\'e}ment, Sanchez-Palencia, Bouyer, and
  Aspect}}]{Bouyer:AndersonBEC:N08}
\bibinfo{author}{\bibfnamefont{J.}~\bibnamefont{Billy}},
  \bibinfo{author}{\bibfnamefont{V.}~\bibnamefont{Josse}},
  \bibinfo{author}{\bibfnamefont{Z.}~\bibnamefont{Zuo}},
  \bibinfo{author}{\bibfnamefont{A.}~\bibnamefont{Bernard}},
  \bibinfo{author}{\bibfnamefont{B.}~\bibnamefont{Hambrecht}},
  \bibinfo{author}{\bibfnamefont{P.}~\bibnamefont{Lugan}},
  \bibinfo{author}{\bibfnamefont{D.}~\bibnamefont{Cl{\'e}ment}},
  \bibinfo{author}{\bibfnamefont{L.}~\bibnamefont{Sanchez-Palencia}},
  \bibinfo{author}{\bibfnamefont{P.}~\bibnamefont{Bouyer}}, \bibnamefont{and}
  \bibinfo{author}{\bibfnamefont{A.}~\bibnamefont{Aspect}},
  \bibinfo{journal}{Nature (London)} \textbf{\bibinfo{volume}{453}},
  \bibinfo{pages}{891} (\bibinfo{year}{2008}).

\bibitem[{\citenamefont{Roati et~al.}(2008)\citenamefont{Roati, d'Errico,
  Fallani, Fattori, Fort, Zaccanti, Modugno, Modugno, and
  Inguscio}}]{Inguscio:AndersonBEC:N08}
\bibinfo{author}{\bibfnamefont{G.}~\bibnamefont{Roati}},
  \bibinfo{author}{\bibfnamefont{C.}~\bibnamefont{d'Errico}},
  \bibinfo{author}{\bibfnamefont{L.}~\bibnamefont{Fallani}},
  \bibinfo{author}{\bibfnamefont{M.}~\bibnamefont{Fattori}},
  \bibinfo{author}{\bibfnamefont{C.}~\bibnamefont{Fort}},
  \bibinfo{author}{\bibfnamefont{M.}~\bibnamefont{Zaccanti}},
  \bibinfo{author}{\bibfnamefont{G.}~\bibnamefont{Modugno}},
  \bibinfo{author}{\bibfnamefont{M.}~\bibnamefont{Modugno}}, \bibnamefont{and}
  \bibinfo{author}{\bibfnamefont{M.}~\bibnamefont{Inguscio}},
  \bibinfo{journal}{Nature (London)} \textbf{\bibinfo{volume}{453}},
  \bibinfo{pages}{895} (\bibinfo{year}{2008}).

\bibitem[{\citenamefont{Deissler et~al.}(2010)\citenamefont{Deissler, Zaccanti,
  Roati, d'Errico, Fattori, Modugno, Modugno, and
  Inguscio}}]{Inguscio:AubryAndreInteractions:NP10}
\bibinfo{author}{\bibfnamefont{B.}~\bibnamefont{Deissler}},
  \bibinfo{author}{\bibfnamefont{M.}~\bibnamefont{Zaccanti}},
  \bibinfo{author}{\bibfnamefont{G.}~\bibnamefont{Roati}},
  \bibinfo{author}{\bibfnamefont{C.}~\bibnamefont{d'Errico}},
  \bibinfo{author}{\bibfnamefont{M.}~\bibnamefont{Fattori}},
  \bibinfo{author}{\bibfnamefont{M.}~\bibnamefont{Modugno}},
  \bibinfo{author}{\bibfnamefont{G.}~\bibnamefont{Modugno}}, \bibnamefont{and}
  \bibinfo{author}{\bibfnamefont{M.}~\bibnamefont{Inguscio}},
  \bibinfo{journal}{Nature Phys.} \textbf{\bibinfo{volume}{6}},
  \bibinfo{pages}{354} (\bibinfo{year}{2010}).

\bibitem[{\citenamefont{Chab{\'e} et~al.}(2008)\citenamefont{Chab{\'e},
  Lemari{\'e}, Gr{\'e}maud, Delande, Szriftgiser, and
  Garreau}}]{AP:Anderson:PRL08}
\bibinfo{author}{\bibfnamefont{J.}~\bibnamefont{Chab{\'e}}},
  \bibinfo{author}{\bibfnamefont{G.}~\bibnamefont{Lemari{\'e}}},
  \bibinfo{author}{\bibfnamefont{B.}~\bibnamefont{Gr{\'e}maud}},
  \bibinfo{author}{\bibfnamefont{D.}~\bibnamefont{Delande}},
  \bibinfo{author}{\bibfnamefont{P.}~\bibnamefont{Szriftgiser}},
  \bibnamefont{and} \bibinfo{author}{\bibfnamefont{J.~C.}
  \bibnamefont{Garreau}}, \bibinfo{journal}{Phys. Rev. Lett.}
  \textbf{\bibinfo{volume}{101}}, \bibinfo{pages}{255702}
  (\bibinfo{year}{2008}).

\bibitem[{\citenamefont{Lemari{\'e} et~al.}(2010)\citenamefont{Lemari{\'e},
  Lignier, Delande, Szriftgiser, and Garreau}}]{AP:AndersonCritic:PRL10}
\bibinfo{author}{\bibfnamefont{G.}~\bibnamefont{Lemari{\'e}}},
  \bibinfo{author}{\bibfnamefont{H.}~\bibnamefont{Lignier}},
  \bibinfo{author}{\bibfnamefont{D.}~\bibnamefont{Delande}},
  \bibinfo{author}{\bibfnamefont{P.}~\bibnamefont{Szriftgiser}},
  \bibnamefont{and} \bibinfo{author}{\bibfnamefont{J.~C.}
  \bibnamefont{Garreau}}, \bibinfo{journal}{Phys. Rev. Lett.}
  \textbf{\bibinfo{volume}{105}}, \bibinfo{pages}{090601}
  (\bibinfo{year}{2010}).

\bibitem[{\citenamefont{Haller et~al.}(2010)\citenamefont{Haller, Hart, Mark,
  Danzl, Reichs{\"o}llner, and N{\"a}gerl}}]{Naegerl:SuperBlochOsc:PRL10}
\bibinfo{author}{\bibfnamefont{E.}~\bibnamefont{Haller}},
  \bibinfo{author}{\bibfnamefont{R.}~\bibnamefont{Hart}},
  \bibinfo{author}{\bibfnamefont{M.~J.} \bibnamefont{Mark}},
  \bibinfo{author}{\bibfnamefont{J.~G.} \bibnamefont{Danzl}},
  \bibinfo{author}{\bibfnamefont{L.}~\bibnamefont{Reichs{\"o}llner}},
  \bibnamefont{and} \bibinfo{author}{\bibfnamefont{H.~C.}
  \bibnamefont{N{\"a}gerl}}, \bibinfo{journal}{Phys. Rev. Lett.}
  \textbf{\bibinfo{volume}{104}}, \bibinfo{pages}{200403}
  (\bibinfo{year}{2010}).

\bibitem[{\citenamefont{Monroe et~al.}(1996)\citenamefont{Monroe, Meekhof,
  King, and Wineland}}]{MonroeWineland:SchrCat:S96}
\bibinfo{author}{\bibfnamefont{C.}~\bibnamefont{Monroe}},
  \bibinfo{author}{\bibfnamefont{D.~M.} \bibnamefont{Meekhof}},
  \bibinfo{author}{\bibfnamefont{B.~E.} \bibnamefont{King}}, \bibnamefont{and}
  \bibinfo{author}{\bibfnamefont{D.~J.} \bibnamefont{Wineland}},
  \bibinfo{journal}{Science} \textbf{\bibinfo{volume}{272}},
  \bibinfo{pages}{1131} (\bibinfo{year}{1996}).

\bibitem[{\citenamefont{Monroe}(2002)}]{Monroe:QIReview:N02}
\bibinfo{author}{\bibfnamefont{C.}~\bibnamefont{Monroe}},
  \bibinfo{journal}{Nature (London)} \textbf{\bibinfo{volume}{416}},
  \bibinfo{pages}{238} (\bibinfo{year}{2002}).

\bibitem[{\citenamefont{Lee et~al.}(2004)\citenamefont{Lee, Villeneuve, Corkum,
  and Shapiro}}]{Shapiro:WavepacketsQI:PRL04}
\bibinfo{author}{\bibfnamefont{K.~F.} \bibnamefont{Lee}},
  \bibinfo{author}{\bibfnamefont{D.~M.} \bibnamefont{Villeneuve}},
  \bibinfo{author}{\bibfnamefont{P.~B.} \bibnamefont{Corkum}},
  \bibnamefont{and} \bibinfo{author}{\bibfnamefont{E.~A.}
  \bibnamefont{Shapiro}}, \bibinfo{journal}{Phys. Rev. Lett.}
  \textbf{\bibinfo{volume}{93}}, \bibinfo{pages}{233601}
  (\bibinfo{year}{2004}).

\bibitem[{\citenamefont{Lengwenus et~al.}(2010)\citenamefont{Lengwenus, Kruse,
  Schlosser, Tichelmann, and Birkl}}]{Lengwenus:CohTransport:PRL10}
\bibinfo{author}{\bibfnamefont{A.}~\bibnamefont{Lengwenus}},
  \bibinfo{author}{\bibfnamefont{J.}~\bibnamefont{Kruse}},
  \bibinfo{author}{\bibfnamefont{M.}~\bibnamefont{Schlosser}},
  \bibinfo{author}{\bibfnamefont{S.}~\bibnamefont{Tichelmann}},
  \bibnamefont{and} \bibinfo{author}{\bibfnamefont{G.}~\bibnamefont{Birkl}},
  \bibinfo{journal}{Phys. Rev. Lett.} \textbf{\bibinfo{volume}{105}},
  \bibinfo{pages}{170502} (\bibinfo{year}{2010}).

\bibitem[{\citenamefont{Ye et~al.}(2008)\citenamefont{Ye, Kimble, and
  Katori}}]{Kimble:QuantumStateMetrol:S08}
\bibinfo{author}{\bibfnamefont{J.}~\bibnamefont{Ye}},
  \bibinfo{author}{\bibfnamefont{H.~J.} \bibnamefont{Kimble}},
  \bibnamefont{and} \bibinfo{author}{\bibfnamefont{H.}~\bibnamefont{Katori}},
  \bibinfo{journal}{Science} \textbf{\bibinfo{volume}{320}},
  \bibinfo{pages}{1734} (\bibinfo{year}{2008}).

\bibitem[{\citenamefont{Ben~Dahan et~al.}(1996)\citenamefont{Ben~Dahan, Peik,
  Reichel, Castin, and Salomon}}]{BenDahan:BlochOsc:PRL96}
\bibinfo{author}{\bibfnamefont{M.}~\bibnamefont{Ben~Dahan}},
  \bibinfo{author}{\bibfnamefont{E.}~\bibnamefont{Peik}},
  \bibinfo{author}{\bibfnamefont{J.}~\bibnamefont{Reichel}},
  \bibinfo{author}{\bibfnamefont{Y.}~\bibnamefont{Castin}}, \bibnamefont{and}
  \bibinfo{author}{\bibfnamefont{C.}~\bibnamefont{Salomon}},
  \bibinfo{journal}{Phys. Rev. Lett.} \textbf{\bibinfo{volume}{76}},
  \bibinfo{pages}{4508} (\bibinfo{year}{1996}).

\bibitem[{\citenamefont{Wilkinson et~al.}(1996)\citenamefont{Wilkinson,
  Bharucha, Madison, Niu, and Raizen}}]{Raizen:WSOptPot:PRL96}
\bibinfo{author}{\bibfnamefont{S.~R.} \bibnamefont{Wilkinson}},
  \bibinfo{author}{\bibfnamefont{C.~F.} \bibnamefont{Bharucha}},
  \bibinfo{author}{\bibfnamefont{K.~W.} \bibnamefont{Madison}},
  \bibinfo{author}{\bibfnamefont{Q.}~\bibnamefont{Niu}}, \bibnamefont{and}
  \bibinfo{author}{\bibfnamefont{M.~G.} \bibnamefont{Raizen}},
  \bibinfo{journal}{Phys. Rev. Lett.} \textbf{\bibinfo{volume}{76}},
  \bibinfo{pages}{4512} (\bibinfo{year}{1996}).

\bibitem[{\citenamefont{Moore et~al.}(1995)\citenamefont{Moore, Robinson,
  Bharucha, Sundaram, and Raizen}}]{Raizen:QKRFirst:PRL95}
\bibinfo{author}{\bibfnamefont{F.~L.} \bibnamefont{Moore}},
  \bibinfo{author}{\bibfnamefont{J.~C.} \bibnamefont{Robinson}},
  \bibinfo{author}{\bibfnamefont{C.~F.} \bibnamefont{Bharucha}},
  \bibinfo{author}{\bibfnamefont{B.}~\bibnamefont{Sundaram}}, \bibnamefont{and}
  \bibinfo{author}{\bibfnamefont{M.~G.} \bibnamefont{Raizen}},
  \bibinfo{journal}{Phys. Rev. Lett.} \textbf{\bibinfo{volume}{75}},
  \bibinfo{pages}{4598} (\bibinfo{year}{1995}).

\bibitem[{\citenamefont{Lignier et~al.}(2005)\citenamefont{Lignier, Chab{\'e},
  Delande, Garreau, and Szriftgiser}}]{AP:Reversibility:PRL05}
\bibinfo{author}{\bibfnamefont{H.}~\bibnamefont{Lignier}},
  \bibinfo{author}{\bibfnamefont{J.}~\bibnamefont{Chab{\'e}}},
  \bibinfo{author}{\bibfnamefont{D.}~\bibnamefont{Delande}},
  \bibinfo{author}{\bibfnamefont{J.~C.} \bibnamefont{Garreau}},
  \bibnamefont{and}
  \bibinfo{author}{\bibfnamefont{P.}~\bibnamefont{Szriftgiser}},
  \bibinfo{journal}{Phys. Rev. Lett.} \textbf{\bibinfo{volume}{95}},
  \bibinfo{pages}{234101} (\bibinfo{year}{2005}).

\bibitem[{\citenamefont{Ringot et~al.}(2000)\citenamefont{Ringot, Szriftgiser,
  Garreau, and Delande}}]{AP:Bicolor:PRL00}
\bibinfo{author}{\bibfnamefont{J.}~\bibnamefont{Ringot}},
  \bibinfo{author}{\bibfnamefont{P.}~\bibnamefont{Szriftgiser}},
  \bibinfo{author}{\bibfnamefont{J.~C.} \bibnamefont{Garreau}},
  \bibnamefont{and} \bibinfo{author}{\bibfnamefont{D.}~\bibnamefont{Delande}},
  \bibinfo{journal}{Phys. Rev. Lett.} \textbf{\bibinfo{volume}{85}},
  \bibinfo{pages}{2741} (\bibinfo{year}{2000}).

\bibitem[{\citenamefont{Talukdar et~al.}(2010)\citenamefont{Talukdar, Shrestha,
  and Summy}}]{Summy:SubFQRes:PRL10}
\bibinfo{author}{\bibfnamefont{I.}~\bibnamefont{Talukdar}},
  \bibinfo{author}{\bibfnamefont{R.}~\bibnamefont{Shrestha}}, \bibnamefont{and}
  \bibinfo{author}{\bibfnamefont{G.~S.} \bibnamefont{Summy}},
  \bibinfo{journal}{Phys. Rev. Lett.} \textbf{\bibinfo{volume}{105}},
  \bibinfo{pages}{054103} (\bibinfo{year}{2010}).

\bibitem[{\citenamefont{Jones et~al.}(2007)\citenamefont{Jones, Goonasekera,
  Meacher, Jonckheere, and Monteiro}}]{Monteiro:DirectedMotionDKR:PRL07}
\bibinfo{author}{\bibfnamefont{P.~H.} \bibnamefont{Jones}},
  \bibinfo{author}{\bibfnamefont{M.}~\bibnamefont{Goonasekera}},
  \bibinfo{author}{\bibfnamefont{D.~R.} \bibnamefont{Meacher}},
  \bibinfo{author}{\bibfnamefont{T.}~\bibnamefont{Jonckheere}},
  \bibnamefont{and} \bibinfo{author}{\bibfnamefont{T.~S.}
  \bibnamefont{Monteiro}}, \bibinfo{journal}{Phys. Rev. Lett.}
  \textbf{\bibinfo{volume}{98}}, \bibinfo{pages}{073002}
  (\bibinfo{year}{2007}).

\bibitem[{\citenamefont{Witthaut et~al.}(2005)\citenamefont{Witthaut, Werder,
  Mossmann, and Korsch}}]{Korsch:BlochOscBEC:PRE05}
\bibinfo{author}{\bibfnamefont{D.}~\bibnamefont{Witthaut}},
  \bibinfo{author}{\bibfnamefont{M.}~\bibnamefont{Werder}},
  \bibinfo{author}{\bibfnamefont{S.}~\bibnamefont{Mossmann}}, \bibnamefont{and}
  \bibinfo{author}{\bibfnamefont{H.~J.} \bibnamefont{Korsch}},
  \bibinfo{journal}{Phys. Rev. E} \textbf{\bibinfo{volume}{71}},
  \bibinfo{pages}{036625} (\bibinfo{year}{2005}).

\bibitem[{\citenamefont{Sadgrove and
  Wimberger}(2009)}]{Sadgrove:TrasportQR:NJP09}
\bibinfo{author}{\bibfnamefont{M.}~\bibnamefont{Sadgrove}} \bibnamefont{and}
  \bibinfo{author}{\bibfnamefont{S.}~\bibnamefont{Wimberger}},
  \bibinfo{journal}{New J. Phys} \textbf{\bibinfo{volume}{11}},
  \bibinfo{pages}{083027} (\bibinfo{year}{2009}).

\bibitem[{\citenamefont{Jaksch and
  Zoller}(2005)}]{Zoller:ControlledCollision:ANP05}
\bibinfo{author}{\bibfnamefont{D.}~\bibnamefont{Jaksch}} \bibnamefont{and}
  \bibinfo{author}{\bibfnamefont{P.}~\bibnamefont{Zoller}},
  \bibinfo{journal}{Ann. Phys.} \textbf{\bibinfo{volume}{315}},
  \bibinfo{pages}{52} (\bibinfo{year}{2005}).

\bibitem[{\citenamefont{Miroshnychenko
  et~al.}(2006)\citenamefont{Miroshnychenko, Alt, Dotsenko, F{\"o}rster,
  Khudaverdyan, Meschede, and
  Rauschenbeutel}}]{Meschede:AtomSortingMachine:N06}
\bibinfo{author}{\bibfnamefont{Y.}~\bibnamefont{Miroshnychenko}},
  \bibinfo{author}{\bibfnamefont{W.}~\bibnamefont{Alt}},
  \bibinfo{author}{\bibfnamefont{I.}~\bibnamefont{Dotsenko}},
  \bibinfo{author}{\bibfnamefont{L.}~\bibnamefont{F{\"o}rster}},
  \bibinfo{author}{\bibfnamefont{M.}~\bibnamefont{Khudaverdyan}},
  \bibinfo{author}{\bibfnamefont{D.}~\bibnamefont{Meschede}}, \bibnamefont{and}
  \bibinfo{author}{\bibfnamefont{D.~S.} \bibnamefont{Rauschenbeutel}},
  \bibinfo{journal}{Nature (London)} \textbf{\bibinfo{volume}{442}},
  \bibinfo{pages}{151} (\bibinfo{year}{2006}).

\bibitem[{\citenamefont{Witthaut et~al.}(2004)\citenamefont{Witthaut, Keck,
  Korsch, and Mossmann}}]{Korsch:BlochOsc2D:NPJ04}
\bibinfo{author}{\bibfnamefont{D.}~\bibnamefont{Witthaut}},
  \bibinfo{author}{\bibfnamefont{F.}~\bibnamefont{Keck}},
  \bibinfo{author}{\bibfnamefont{H.~J.} \bibnamefont{Korsch}},
  \bibnamefont{and} \bibinfo{author}{\bibfnamefont{S.}~\bibnamefont{Mossmann}},
  \bibinfo{journal}{New J. Phys} \textbf{\bibinfo{volume}{6}},
  \bibinfo{pages}{41} (\bibinfo{year}{2004}).

\bibitem[{\citenamefont{Zenesini et~al.}(2009)\citenamefont{Zenesini, Lignier,
  Tayebirad, Radogostowicz, Ciampini, Mannella, Wimberger, Morsch, and
  Arimondo}}]{Arimondo:LandauZener:PRL09}
\bibinfo{author}{\bibfnamefont{A.}~\bibnamefont{Zenesini}},
  \bibinfo{author}{\bibfnamefont{H.}~\bibnamefont{Lignier}},
  \bibinfo{author}{\bibfnamefont{G.}~\bibnamefont{Tayebirad}},
  \bibinfo{author}{\bibfnamefont{J.}~\bibnamefont{Radogostowicz}},
  \bibinfo{author}{\bibfnamefont{D.}~\bibnamefont{Ciampini}},
  \bibinfo{author}{\bibfnamefont{R.}~\bibnamefont{Mannella}},
  \bibinfo{author}{\bibfnamefont{S.}~\bibnamefont{Wimberger}},
  \bibinfo{author}{\bibfnamefont{O.}~\bibnamefont{Morsch}}, \bibnamefont{and}
  \bibinfo{author}{\bibfnamefont{E.}~\bibnamefont{Arimondo}},
  \bibinfo{journal}{Phys. Rev. Lett.} \textbf{\bibinfo{volume}{103}},
  \bibinfo{pages}{090403} (\bibinfo{year}{2009}).

\bibitem[{\citenamefont{Wannier}(1960)}]{Wannier:WS:PR60}
\bibinfo{author}{\bibfnamefont{G.~H.} \bibnamefont{Wannier}},
  \bibinfo{journal}{Phys. Rev.} \textbf{\bibinfo{volume}{117}},
  \bibinfo{pages}{432} (\bibinfo{year}{1960}).

\bibitem[{\citenamefont{Thommen et~al.}(2002)\citenamefont{Thommen, Garreau,
  and Zehnl{\'e}}}]{AP:WannierStark:PRA02}
\bibinfo{author}{\bibfnamefont{Q.}~\bibnamefont{Thommen}},
  \bibinfo{author}{\bibfnamefont{J.~C.} \bibnamefont{Garreau}},
  \bibnamefont{and}
  \bibinfo{author}{\bibfnamefont{V.}~\bibnamefont{Zehnl{\'e}}},
  \bibinfo{journal}{Phys. Rev. A} \textbf{\bibinfo{volume}{65}},
  \bibinfo{pages}{053406} (\bibinfo{year}{2002}).

\bibitem[{\citenamefont{Haroutyunyan and
  Nienhuis}(2001)}]{Nienhuis:CoherentDyn:PRA01}
\bibinfo{author}{\bibfnamefont{H.~L.} \bibnamefont{Haroutyunyan}}
  \bibnamefont{and} \bibinfo{author}{\bibfnamefont{G.}~\bibnamefont{Nienhuis}},
  \bibinfo{journal}{Phys. Rev. A} \textbf{\bibinfo{volume}{64}},
  \bibinfo{pages}{033424} (\bibinfo{year}{2001}).

\bibitem[{\citenamefont{Gl{\"u}ck et~al.}(2000)\citenamefont{Gl{\"u}ck, Hankel,
  Kolovsky, and Korsch}}]{Korsh:WSOptLatt:PRA00}
\bibinfo{author}{\bibfnamefont{M.}~\bibnamefont{Gl{\"u}ck}},
  \bibinfo{author}{\bibfnamefont{M.}~\bibnamefont{Hankel}},
  \bibinfo{author}{\bibfnamefont{A.~R.} \bibnamefont{Kolovsky}},
  \bibnamefont{and} \bibinfo{author}{\bibfnamefont{H.~J.}
  \bibnamefont{Korsch}}, \bibinfo{journal}{Phys. Rev. A}
  \textbf{\bibinfo{volume}{61}}, \bibinfo{pages}{061402(R)}
  (\bibinfo{year}{2000}).

\bibitem[{\citenamefont{Thommen et~al.}(2004)\citenamefont{Thommen, Garreau,
  and Zehnl{\'e}}}]{AP:QuantumDyn:JOBQSO04}
\bibinfo{author}{\bibfnamefont{Q.}~\bibnamefont{Thommen}},
  \bibinfo{author}{\bibfnamefont{J.~C.} \bibnamefont{Garreau}},
  \bibnamefont{and}
  \bibinfo{author}{\bibfnamefont{V.}~\bibnamefont{Zehnl{\'e}}},
  \bibinfo{journal}{J. Opt. B: Quantum Semiclass. Opt.}
  \textbf{\bibinfo{volume}{6}}, \bibinfo{pages}{301} (\bibinfo{year}{2004}).

\bibitem[{\citenamefont{Carlo et~al.}(2006)\citenamefont{Carlo, Benenti,
  Casati, Wimberger, Morsch, Mannella, and
  Arimondo}}]{Arimondo:RatchetOptLat:PRA06}
\bibinfo{author}{\bibfnamefont{G.~G.} \bibnamefont{Carlo}},
  \bibinfo{author}{\bibfnamefont{G.}~\bibnamefont{Benenti}},
  \bibinfo{author}{\bibfnamefont{G.}~\bibnamefont{Casati}},
  \bibinfo{author}{\bibfnamefont{S.}~\bibnamefont{Wimberger}},
  \bibinfo{author}{\bibfnamefont{O.}~\bibnamefont{Morsch}},
  \bibinfo{author}{\bibfnamefont{R.}~\bibnamefont{Mannella}}, \bibnamefont{and}
  \bibinfo{author}{\bibfnamefont{E.}~\bibnamefont{Arimondo}},
  \bibinfo{journal}{Phys. Rev. A} \textbf{\bibinfo{volume}{74}},
  \bibinfo{pages}{033617} (\bibinfo{year}{2006}).

\end{thebibliography}

\end{document}